\documentclass[twocolumn]{aastex6}

\newcommand{\J}{\mbox{J0815+0939}}
\newcommand{\B}{\mbox{B1839-04~}}

\submitted{}
\shorttitle{Bi-drifting subpulse phenomena in pulsars}
\shortauthors{Szary \& van Leeuwen}

\begin{document}

\title{On the origin of the bi-drifting subpulse phenomenon in pulsars }
\author{Andrzej Szary\altaffilmark{1, 2} and Joeri van Leeuwen\altaffilmark{1,3}}
\email{szary@astron.nl}

\altaffiltext{1}{ASTRON, the Netherlands Institute for Radio Astronomy, Postbus 2, 7990 AA, Dwingeloo, The Netherlands}
\altaffiltext{2}{Janusz Gil Institute of Astronomy, University of Zielona G\'ora, Lubuska 2, 65-265 Zielona G\'ora, Poland}
\altaffiltext{3}{Anton Pannekoek Institute for Astronomy, University of Amsterdam, Science Park 904, 1098 XH Amsterdam, Netherlands}

\keywords{pulsars: general --- pulsars: individual ({J0815+0939})
\newline \newline \hspace*{-2cm} \it This version of the article incorporates corrections mentioned in the Erratum (2019, ApJ, 878,163).}

\begin{abstract}
The unique and highly unusual drift feature reported for PSR J0815+0939, wherein one component's subpulses drift in the direction opposite of the general trend, is a veritable challenge to pulsar theory.
In this paper, we observationally quantify the drift direction throughout its profile, and find that the second component is the only one that exhibits "bi-drifting", meaning that only second component moves in the direction opposite of the others.
We here present a model that shows that the observed bi-drifting phenomenon follows from the insight that the discharging regions, i.e. sparks, do not rotate around the magnetic axis per se, but rather around the point of electric potential extremum at the polar cap: minimum in the pulsar case ($\mathbf{\Omega} \cdot \mathbf{B} < 0$) and maximum in the antipulsar case ($\mathbf{\Omega} \cdot \mathbf{B} > 0$).
We show that a purely dipolar surface magnetic field cannot exhibit bi-drifting behaviour, though certain non-dipolar configurations can.
We can distinguish two types of solutions, with relatively low ($\sim 10^{12} \,{\rm G}$) and high ($\sim 10^{14} \,{\rm G}$) surface magnetic fields.
Depending on the strength of the surface magnetic field, the radius of the curvature of magnetic field lines ranges from $10^{5} \,{\rm cm}$ to $10^{7}\,{\rm cm}$.
Pulsar \J~allows us to gain an understanding of the polar-cap conditions essential for plasma generation processes in the inner acceleration region, by linking the observed subpulse shift to the underlying spark motion.
\end{abstract}

\section{Introduction}

Drifting subpulses, the intriguing if not baffling trend of pulsar emission features to steadily march through subsequent pulses, were modeled by \cite{1975_Ruderman} as localized pair cascades, which are discharges of small regions over the polar cap.
Such sparks produce plasma columns that stream into the magnetosphere, where they produce the observed radio emission \citep[see][for more details]{1998_Asseo,2000_Melikidze,2015_Szary}.
In the natural force-free state, with a Lorentz force of zero, the magnetosphere is filled with charged particles with the so-called Goldreich-Julian density $\rho_{\rm GJ}=-{\bf \Omega}\cdot{\bf B} / (2 \pi c)$.
Then the electric field $\bf E_{\perp}$ perpendicular to the magnetic field $\bf B$ is
\begin{equation}
    {\bf E_{ \perp} } = -\frac{1}{c} ({\bf \Omega} \times {\bf r}) \times {\bf B}, 
    \label{eq:ff}
\end{equation}
where $\bf \Omega$ is the angular velocity, $\bf r$ is the location vector, and $c$ is the speed of light.
In this force-free state, particles co-rotate with the star with velocity \mbox{${\bf v_{\rm cor}} = {c\left ( \bf {E}_{\perp} \times B \right)}/{ B^2}$}.

By definition, in the acceleration region, the density of charges is less than the co-rotational charge density.
Plasma in this region is accelerated along the magnetic field and moves perpendicular to the magnetic field with velocity
\begin{equation}
 {\bf v} = \frac{c(\bf \widetilde{E}_{\perp} \times B)}{B^2},
\end{equation}
where ${\bf  \widetilde{E}_{\perp}} $ is the electric field in the plasma starved region perpendicular to the magnetic field.
Note that in the plasma starved region ${  \widetilde{\bf E}_\perp} \neq {\bf E_{\perp}}$; furthermore, the value of an electric field in some areas of these regions can be higher than in the co-rotating region (see Section \ref{sec:theoretical_background}).

Drift of subpulses around the magnetic axis was first introduced by \cite{1975_Ruderman}.
It turned out that the carousel-like rotation of sparks around the magnetic axis can explain a variety of pulsar data \citep[see, e.g.,][]{2000_Gil,2001_Gil, 2006_Weltevrede, 2007_Weltevrede, 2007_Herfindal, 2008_Rankin, 2009_Herfindal, 2014_Rankin}.
Furthermore, it was shown by \citet{1993ppm..book.....G} and \citet{2010mfca.book.....B} that a potential difference between the central and periphery domains over the acceleration region leads to the rotation of the plasma around the magnetic axis.
This strengthened confidence in the validity of the carousel model for oblique pulsars.

The plasma responsible for radio emission is generated and accelerated in the inner acceleration region (IAR) just above the polar cap, while the radio emission is generated at much higher altitudes.
In \cite{2012_Leeuwen}, the drift velocity was shown to depend on the variation of the potential drop over the polar cap.
We use this prescription to characterize the plasma rotation.
%We show that this transition from the polar cap region to the emission region influences the observed drift characteristics.

In pulsars with multiple components, subpulses generally drift in the same  direction. 
There are only two pulsars that exhibit "bi-drifting", in which the emission feature motion between two components is antipodal: \J~\citep{2005_Champion} and \B \citep{2006_Weltevrede}.
One early explanation for the two-direction drift in PSR \J~was proposed by \cite{2004_Qiao}, requiring rotation of sparks around the magnetic axis and the coexistence of an inner annular gap \citep{2004_Qiao_b}, plus the conventional inner gap \citep{1975_Ruderman}.
Furthermore, the proposed model requires the two carousels to contain different numbers of sparks, to circulate with different speeds in opposite directions, and to result in the same repetition time of the drift pulse pattern for all components.
As corroborated  by \cite{2016_Weltevrede} we find this combination highly unlikely, especially since both bi-drifting pulsars share this property.
\cite{2017_Wright} propose an alternative explanation.
There is an elliptical, tilted, and extremely eccentric motion of emitting regions around the magnetic axis that can account for the observed bi-drifting subpulses.

In this paper we present a detailed analysis of archival data of PSR \J~ and reveal the systematic drift of the first component. 
We show that we can explain the bi-drifting phenomenon of PSR \J~ in the framework of the \citet{2012_Leeuwen} model.

\section{Observations and Data}

\subsection{Data Taking}

\begin{figure}[ht]
  \centerline{\includegraphics[width=5.5cm]{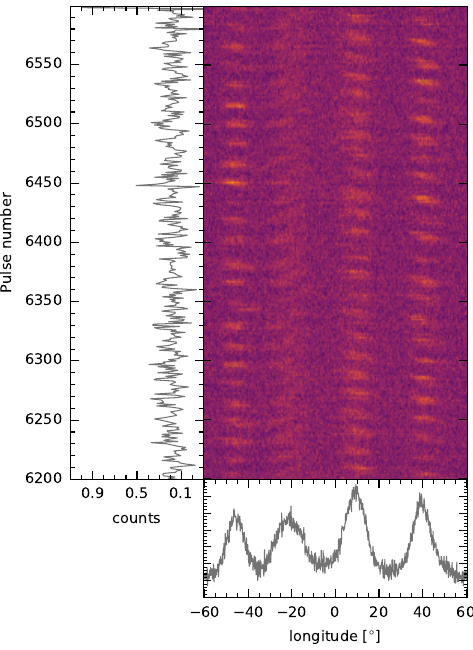}}
    \caption{Single pulses in \J.
    An intensity plot of 400 single pulses with stable $P_3$ is shown in the main panel. The bottom panel shows the integrated profile for this sequence. Variations in the single-pulse intensity are plotted in the left. The data allows for clear distinction between single pulses and the interpretation of the single-pulse patterns.
    \label{fig:single_obs}
    }
\end{figure}

The \J~ data presented in this paper were recorded with the William E. Gordon radio telescope at Arecibo on 2004 June 05.
Over a total time of 5800\,s, the Wideband Arecibo Pulsar Processor (WAPP) recorded 25.9\,MHz of bandwidth around 339\,MHz, at 256\,$\mu$s time resolution, in total intensity Stokes I.
These 9000 pulses were dedispersed at 52.7\,pc cm$^{-3}$ \citep{2005_Champion} and then were folded at the period that maximized the signal-to-noise ratio (S/N) of the four-component peaks. 
The 10\,K/Jy sensitivity provided by the 305 m dish and the 327-MHz Gregorian ("P Band") receiver enables clearly distinct single pulses, as seen in Figure~\ref{fig:single_obs}.

%single_1.pdf

\subsection{Drift Characteristics} \label{sec:drift_characteristics}

    \begin{table*}[t!]
        \caption{Drift parameters of PSR \J }
        \begin{center} 
            \begin{tabular}{lcccc}
                \hline
                \hline
                Component: & I & II & III& IV \\
                Longitude: & 
$-56^{\circ}..-32^{\circ}$ &
$-32^{\circ}..-7^{\circ}$ &
$-7^{\circ}..26^{\circ}$ &
$26^{\circ}..56^{\circ}$ \\

                Drift rate: $\left ( \deg s^{-1} \right )$  &
$ -0.5 \pm  0.1$ &
$  1.11 \pm  0.04$ &
$ -0.59 \pm  0.07$ &
$ -0.87 \pm  0.06$
\\
                \hline
            \end{tabular}
        \end{center}
        \label{tab:drift}
        {\bf Notes.} The standard errors of linear regression are quoted.
    \end{table*}

\begin{figure}[t]
    \begin{center}
        \includegraphics[width=5.5cm]{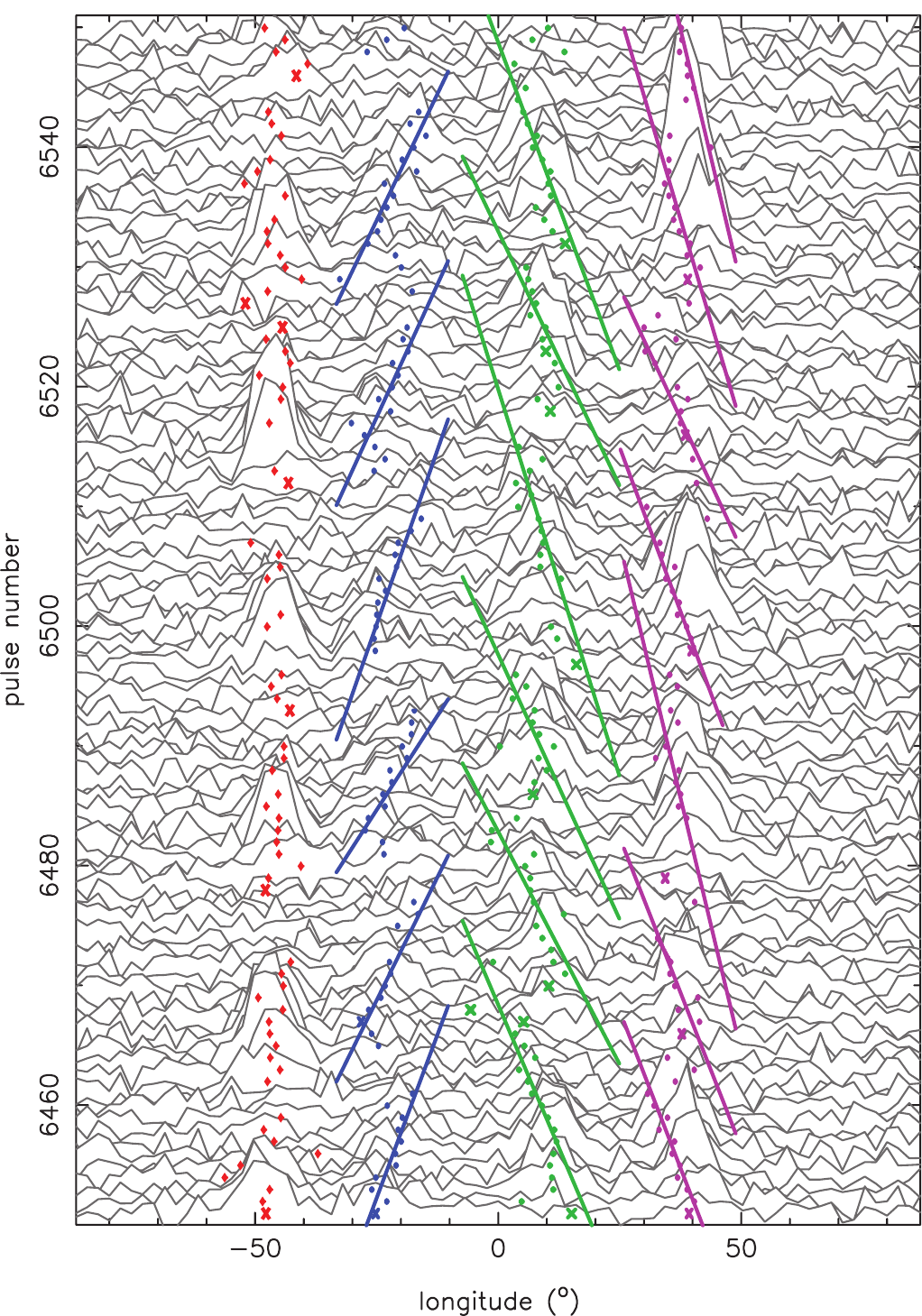}
    \end{center}
    \caption{
        100-pulse sequence with high signal-to-noise. In this figure, data are down sampled by a factor 16 to more clearly display the subpulse locations.
        The single pulses are fitted within each of the four components, and the subsequent drift bands are in the trailing three.
    \label{fig:stacked_fitted}
    }
\end{figure}

From Figure~\ref{fig:single_obs}, four components in longitude can be identified, and there is clear subpulse modulation in intensity and phase within each of these components.
Equally visible by eye is the repetition of the drift pulse pattern; this periodicity $P_3$ is about $\sim$17 pulse periods. 
To quantify this behavior, we aim to abstract the single-pulse trains through fits, and to then model the repeating subpulse pattern visible in Figure~\ref{fig:single_obs}.
Because there is little indication of interaction between components, for fitting purposes, these were treated separately. 
Component boundaries were defined to be at the local minima of the integrated pulse profile.
In that manner, the components were defined as the longitude ranges listed in Table~\ref{tab:drift}, on the scale shown in Figure~\ref{fig:stacked_fitted}.
Within each component, the component subpulse energy was defined as the sum of the comprised samples. Subpulses with low-energy outliers were marked as subpulse "nulls". 
First, one Gaussian was fit within each of the four-component windows. 
The paths that these subpulses trace in time were next fit with straight lines \citep{lkr+02}, ending when the subpulses null (see Figure~\ref{fig:stacked_fitted}).
This was first done for all single pulses, but at a limited signal-to-noise ratio.

%stackeddata-v20.pdf

%lrfs_1
\begin{figure}[t!]
        \centerline{\includegraphics[width=5.5cm]{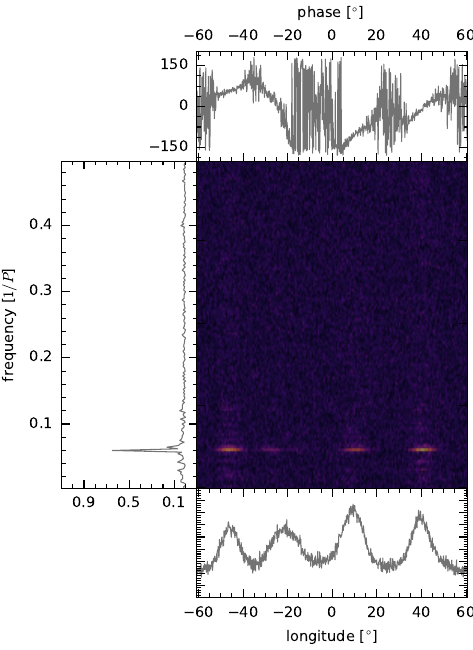}}
        \caption{LRFS plot for the single-pulse signal presented in Figure~\ref{fig:single_obs}.
        The left panel shows the frequency domain in terms of pulsar period.
        The top panel shows the phase variation across the pulse window at the peak amplitude $f=0.0602 \pm 0.0011/P$ (which corresponds to $P_3=16.62 \pm 0.31 P$).
        The bottom panel shows the average profile. 
        }
    \label{fig:lrfs_data}
\end{figure}

%folded_1
\begin{figure}[t!]
    \begin{center}
        \includegraphics[width=6.0cm]{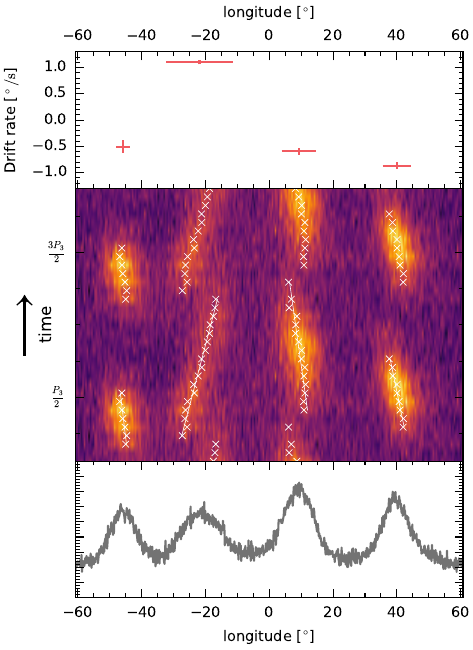}
    \end{center}
    \caption{Average driftband profile, obtained by folding the single-pulse signal presented in Figure~\ref{fig:single_obs} with $P_3=16.62 \, P$ (middle panel, shown twice for clarity).
    The top panel shows the measured drift rate, while the bottom panel shows the integrated pulse profile.}
    \label{fig:folded_data}
\end{figure}

In order to determine $P_3$ with higher precision, we employed fluctuation spectral analysis.
These longitude resolved fluctuation spectra \citep[LRFS,][]{1970_Backer}   involve discrete Fourier transforms of consecutive pulses along each longitude.
The LRFS  find periodicities along pulse longitudes, and explore how the drifting feature varies across the pulse window.
When implemented using a Fourier transform, it produces complex numbers, which can be separated into two parts. The absolute value for a given frequency represents the amount of that frequency in the signal, while the complex component is the phase offset in that frequency.
We thus utilize the angle of the complex argument to study the subpulse motion throughout the pulse window.
In Figure~\ref{fig:lrfs_data} we show the LRFS analysis of the single pulses shown in Figure~\ref{fig:single_obs}.
To determine the frequency peak in the fluctuation spectra, we used the {\it PeakUtils} package. %\footnote{https://pypi.org/project/PeakUtils/}.
Fitting a Gaussian near the major peak in the fluctuation spectra results in frequency $f=0.0602\pm 0.0011 /P$ which corresponds to $P_3=16.62 \pm 0.31 P$ (see the left panel in Figure~\ref{fig:lrfs_data}).
In the top panel of Figure~\ref{fig:lrfs_data} we show the phase variation across the pulse window at the peak amplitude $f=0.0602/P$. 
The slope of the phase variation incorporates information about the direction of the subpulse drift.
We find that only the second component moves in the direction reverse from  the three remaining components (I, III, and IV).
In Figure~\ref{fig:folded_data} we show the folded profile of the sequence of 400 pulses presented in Figure~\ref{fig:single_obs}.
The profile was folded using $P_3=16.62\,P$.

The resulting average drift bands were then fit in the same manner as the individual pulses.
The resulting drift rates are shown in the top panel of Figure~\ref{fig:folded_data} and are listed in Table~\ref{tab:drift}.
The increased signal-to-noise now also allows for the first  detection of a drift rate in component I. 
As can be seen from Figure~\ref{fig:folded_data}, it moves parallel to  components III and IV.
Clearly component II is the only band whose direction is contrary.

In this $P_3$-folded sequence, subpulses and subpulse nulls can be confidently segregated using the method described in \citet{jl04}. The nulling sections between the four components do not occur simultaneously, showing that in this source, such nulling is not a sign of the complete shut-off of the entire pulsar emission mechanism, as appears to be the case in pulsars where the complete pulse profile, including all components, disappears. 
In {\J}, rather, the subpulse nulls are periodic and part of the entire $P_3$ modulation cycle. 

\section{The model}

\subsection{Theoretical background} \label{sec:theoretical_background}

\begin{figure*}[thb]
    \begin{center}
        \includegraphics[width=14cm]{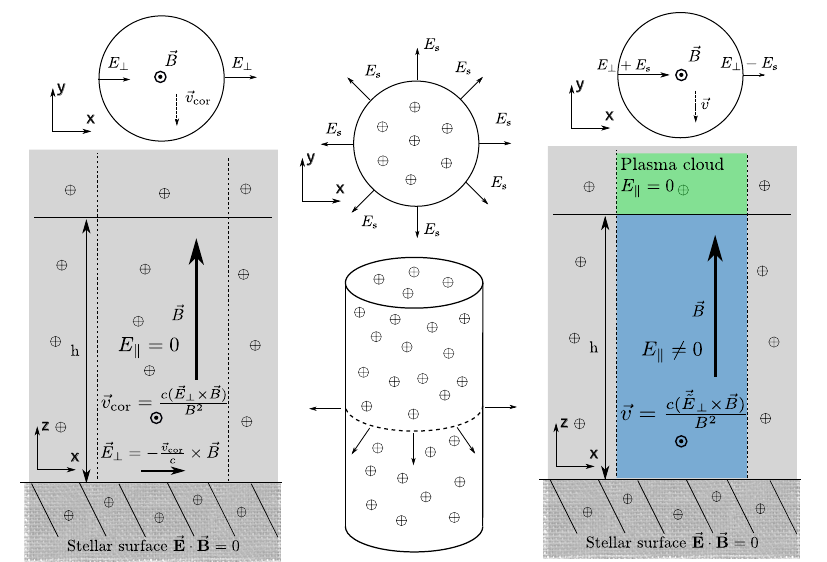}
    \end{center}
    \caption{Electric and magnetic fields within a region of spark formation with screened (the left panel) and unscreened (the right panel) accelerating potential. 
    The middle panel shows the electric field of the plasma column with the co-rotational density.
    }
    \label{fig:spark_region}
\end{figure*}

\begin{figure*}[htb]
    \begin{center}
        \includegraphics[width=13.5cm]{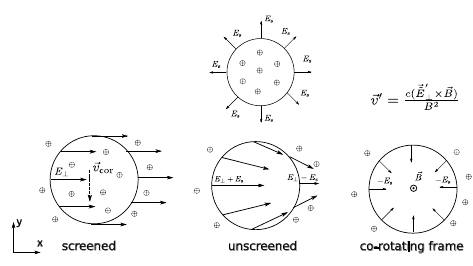}
    \end{center}
    \caption{Electric field in the spark-forming region with screened (the left panel) and unscreened (the middle panel) accelerating potential. 
    The right panel shows the electric field in the co-rotating frame of reference.
    }
    \label{fig:spark_field}
\end{figure*}

\begin{figure}[t!]
    \begin{center}
        \includegraphics[width=8.5cm]{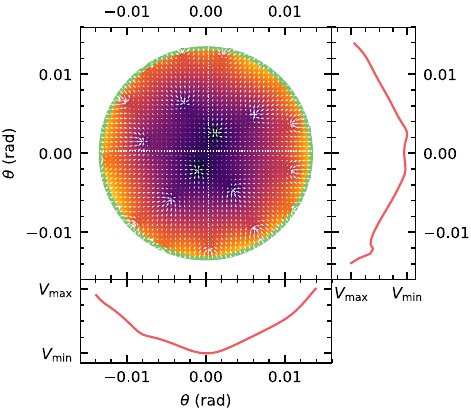}
    \end{center}
    \caption{Electric field (the white arrows) across the polar cap for random distribution of sparks. The color map corresponds to the electric potential. The right and bottom panels show the electric potential for vertical and horizontal cuts throughout the polar cap.
    }
    \label{fig:electric_field}
\end{figure}

\begin{figure}[t!]
    \begin{center}
        \includegraphics[width=7.5cm]{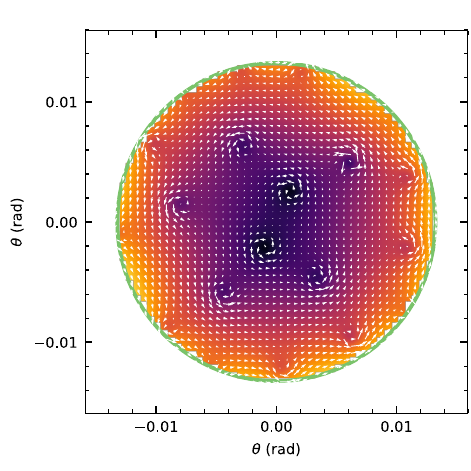}
    \end{center}
    \caption{Plasma velocity (the white arrows) across the polar cap for a random distribution of sparks. The color map corresponds to the electric potential.
    }
    \label{fig:drift_velocity}
\end{figure}

The physics governing the plasma generation in the polar cap region is a major remaining piece of the pulsar puzzle that is not completely solved.
In our studies, we adopt the general approach of \cite{1975_Ruderman}, where sparks form in regions of local extrema of the electrical potential, minima for the pulsar case ($\mathbf{\Omega} \cdot \mathbf{B} < 0$) with a net positive charge at the polar cap, and maxima in the antipulsar case ($\mathbf{\Omega} \cdot \mathbf{B} > 0$) with a net negative charge at the polar cap.
In Figure \ref{fig:spark_region}, we consider a single isolated spark with a cylindrical shape.
At the time when the spark-forming region is filled with plasma with the co-rotational charge density, it co-rotates with the star (see the left panel in Figure \ref{fig:spark_region}).
An accelerating potential emerges when charged particles flow out and leave this region.
The electric field due to the charge enclosed in the spark, $\bf E_{\rm s}$, is presented in the middle panel in Figure \ref{fig:spark_region}.
Note that the lack of electric charges affects the electric field both inside and outside of the region (see the right panel in Figure \ref{fig:spark_region}).
To study the drift of plasma with respect to the co-rotating magnetosphere, we introduce the co-rotating frame of reference.
Due to the Lorentz transformation, the electric field at the boundary of the spark forming region $ \widetilde{\bf  E}^{^{\prime}}_{\perp} = {\bf E_{\perp}} + c^{-1}{{\bf v}_{\rm cor}} \times {\bf B}  - {{\bf E}_{\rm s}} = - {{\bf E}_{\rm s}}$ (see the right panel in Figure \ref{fig:spark_field}).
Such an electric field introduces circulation of plasma around the point of minimum potential. % TODO
Thus, during the discharge in the spark region, the generated plasma should not exhibit any systematic drift with respect to the co-rotating magnetosphere. 
However, the electric potential growing during the discharge also influences the electric field beyond the spark-forming region.
To describe the electric potential of a single spark, we use $V'\propto\ln(r_{\rm s})$, where $r_{\rm s}$ is the distance from the spark center, thus the electric field ${\bf E'}_{\rm s}=-\nabla V'\propto r_{\rm s}^{-1}$.
To study the influence of spark formation on the electric field in the polar cap region, we assume a random distribution of sparks.
In Figure \ref{fig:electric_field} we show the electric field due to randomly distributed sparks across the polar cap.
As the resulting electric potential is a consequence of all forming sparks, its value is lowest toward the center of the polar cap and increases as we approach the polar cap boundary.
Assuming a dipolar configuration of the magnetic field, we estimate its value given in the spherical coordinates to be ${\bf B}=(B_{\rm d}R^{3}r^{-3}\cos{\theta}, B_{\rm d}R^{3}r^{-3}\sin{\theta}, 0)$, where $B_{\rm d}=2\times 10^{12} (P \dot{P}_{-15})^{0.5} \,{\rm G}$ is the polar magnetic field strength, $\dot{P}_{-15}=\dot{P}/10^{-15}$ is the period derivative, $R$ is the neutron star radius, $r$ is the radial distance, and $\theta$ is the polar angle (z-axis coincides with the magnetic axis).
Then, the plasma velocity in the co-rotating frame of reference can be calculated as ${\bf v}' = c(\widetilde{\bf  E}^{^{\prime}}_{\perp} \times {\bf B})B^{-2}$, where $\widetilde{\bf E}_{\perp}^{^{\prime}}$ is the electric field perpendicular to the magnetic field in the co-rotating frame.
Figure \ref{fig:drift_velocity} shows the plasma velocity at the polar cap, for a random distribution of sparks.
As expected, the plasma circulates around the local potential minima.
However, since the electric field between sparks is influenced by all neighboring sparks, an extra circulation of plasma around the global potential minimum at the polar cap occurs.
Note that we have not imposed any additional conditions on the potential variation across the polar cap -- circular-like plasma rotation follows naturally from a random spark distribution.
For the pulsar geometry, the potential is lowest near the center of the polar cap and increases as we approach the polar cap boundary (see the top and left panels in Figure \ref{fig:electric_field}).

Explaining why local regions with low electric potential, where the sparks form, and regions with high potential, between sparks, continue to exist is beyond the scope of this paper \citep[but see, e.g.,][]{2012_Lyubarsky}. 
What we note here is that the observed stability of subpulse structures suggests that the pattern of discharging regions at the polar cap is also stable.
Thus, to study the drift of this stable pattern, we use the global variation of electric potential across the polar cap.

% do not move around!?
% drift_direction.pdf

\clearpage

%\clearpage

\subsection{Neutron star setup}

The magnetic field in our simulation is calculated using the model presented in \cite{2002_Gil}.
It is a result of a global dipole anchored in the center of the star, plus crust-anchored small-scale anomalies.
Such a configuration results in non-dipolar magnetic field at the surface, but the global dipole starts to dominate at a relatively low distance from the star, where the magnetic-field configuration is indistinguishable from a pure dipole (see Figure 3 in \citealt{2015_Szary}).
The coordinate system is chosen such that the z-axis is aligned with the magnetic axis.
Thus, the global field is characterized only by a dipole moment $B_d$, while the surface anomalies are described by a dipole moment $\mathbf{B_{\rm m}}$ and an anomaly location $\mathbf{r_{\rm m}}$.
If not stated otherwise, spherical coordinates of vectors are used.
Note that $B_{\rm m}\ll B_{\rm d}$ to produce a non-dipolar field in the IAR, but a dipolar field at the emission height.
The geometry of a neutron star is further defined by the inclination angle $\alpha$, the opening angle $\rho$, and the impact factor $\beta$.

% ./ns_field.py -m 3 -s -d 401 -f par_0.json -n drift_10 -a central_0 -p 14  (lines2.svg)
% ./ns_field.py -m 13 -s -d 400 -f par_1.json -n drift_10 -a central_0 -p 35 37 38
% simulation
\begin{figure*}
    \begin{center}
        \includegraphics[height=19cm]{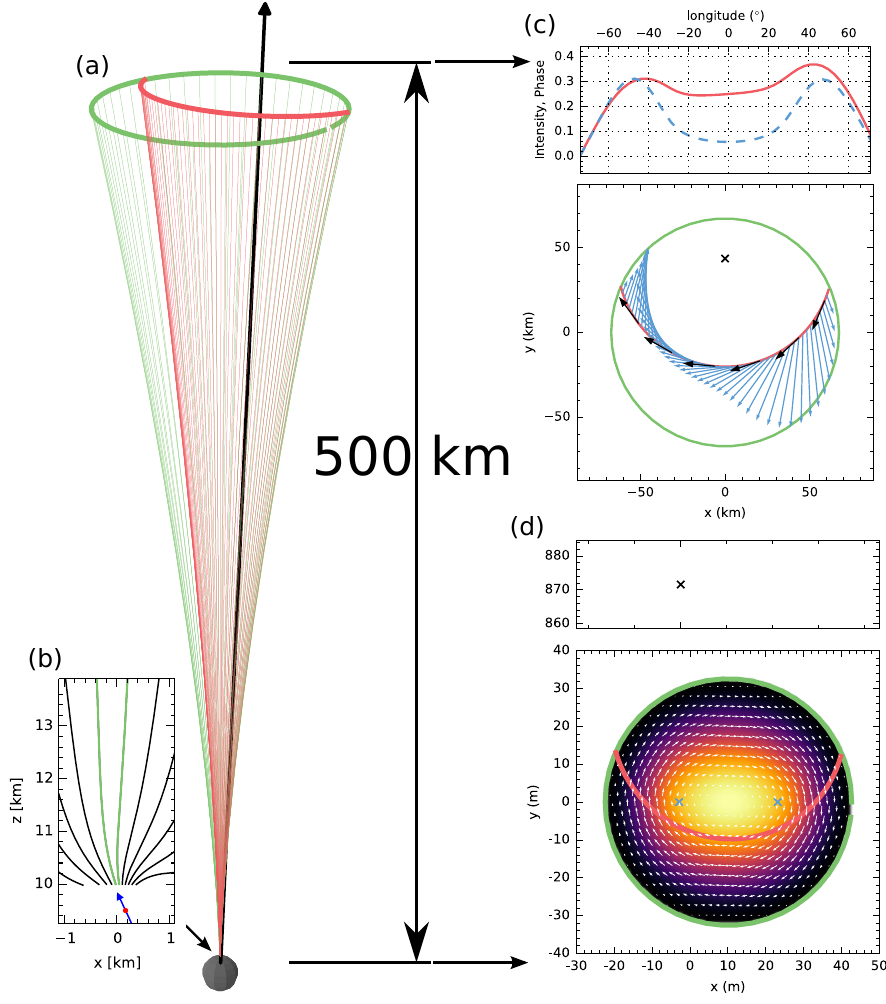}
    \end{center}
    \caption{Results of the calculation using the following parameters:  $\alpha=5^{\circ}$, $\beta / \rho = 0.2$.
    Panel (a): the last open magnetic field lines are drawn in green, while the magnetic lines connected to the observer's line of sight are red. 
    Panel (b): configuration of surface magnetic field calculated using one crust-anchored anomaly (blue arrow) located at ${\bf r_{\rm m}}=\left(0.95R,1\,^{\circ},0\,^{\circ}\right)$ with the dipole moment ${\bf B_{\rm m}} = \left(5\times10^{-3} B_{\rm d},\,25^{\circ},\,180^{\circ}\right)$.
    Panel (c): the green circle is a rim of the open field line region at the emission height, while the red line  corresponds to the observer's line of sight.
    The blue arrows show the drift direction, while the black arrows correspond to the co-rotation direction.
    The top panel shows components of the drift vector along and across the line of sight that are responsible for the phase (red solid line) and intensity (blue dashed line) variation across the pulse window.
    Panel (d): the green circle is a rim of the open field line region at the polar cap, while the red line corresponds to the plasma line (see the text for a description).
    The blue crosses mark locations of potential maxima, while the color map corresponds to the final electric potential.
    The white arrows show the drift direction.
    In both panels (c) and (d), the black cross is the location of the rotation axis at a given height.}
    \label{fig:simulation}
\end{figure*}

\subsection{Drift characteristics}

Once the neutron star configuration is set up, we model the drift characteristics in four steps, which are explained here with the help of Figure \ref{fig:simulation}. 
Panel (a) in Figure~\ref{fig:simulation} shows the magnetic field lines in the open field line region calculated from the neutron star surface up to the emission region at  500\,km.
Panel (b) shows the vertical cut through the structure of the field lines near the stellar surface.
Panel (c) represents the condition in the emission region (horizontal view), while panel (d) represents the condition at the neutron star surface (horizontal view).

(I) We first find the last open field lines in the emission region (the green circle in panel (c)). By tracing these down to the stellar surface (green circle in panel (d)) we outline the actual location of the polar cap.

(II) The second step consists of finding the line of sight at the emission height (the red path in panel (c)), and tracing the open field lines connected to this region down to the stellar surface (the red path in panel (d)).
The red path at the emission region (panel (c)) represents the pulse window, and any subpulse drifting is seen as relative motion of the subpulses with respect to this path.
The sparks are produced just above the polar cap in the IAR, where the charge density is lower than the co-rotation density.
The motion of the outflowing plasma in the IAR is connected with the motion of subpulses in the emission region.
Thus, hereafter, the imprint of the line of sight on the polar cap, which marks the region where the plasma responsible for radio emission is generated, is called the plasma line.
The observed subpulse drift is a consequence of the motion of the spark-forming regions.
The theoretical arguments presented in Section \ref{sec:theoretical_background} suggest that the sparks rotate in a circular-like fashion around the point of potential extremum at the polar cap.
This point can be offset from the magnetic axis.
%Note that during a discharge plasma inside a spark circulates around a local potential maximum (center of a spark), while the plasma drift between spark forming regions is defined by the global potential maximum at the polar cap.

(III) Third, we determine the drift direction, from the variation of the polar-cap electric potential.
To explain the drift properties of PSR \J~ we use the anti-pulsar geometry with $\mathbf{\Omega} \cdot \mathbf{B} > 0$ (see Section \ref{sec:results}), thus the drift direction is determined by maximum of electric potential at the polar cap.  
This potential in the co-rotating frame is modeled as $V'\propto V_{0} \cos{\left( \pi r_{\rm c}/r_{\rm b} \right )}$, where $V_0$ is the potential amplitude, $r_{\rm c}$ is the distance from the location of the potential maximum, and $r_{\rm b}$ is the distance to the polar cap boundary.
Such a simple sinusoidal description of an electric potential not only allows us to define the potential maximum, but also guarantees that the electric field at the polar cap boundary is zero.
To reproduce more complicated potentials, we also consider models with multiple potential maxima (see panel (d) in Figure \ref{fig:simulation}).
Now knowing the electric potential, and thus the electric field $\widetilde{E}^{^{\prime}}_{\perp}=-\nabla V^{\prime}$, we determine the drift direction as described in Section \ref{sec:theoretical_background}.

(IV) The fourth step involves finding of the drift characteristics.
In panel (c) in Figure~\ref{fig:simulation} the blue arrows correspond to the direction of subpulse motion ($\rm{\bf\hat{s}}$), while the black arrows correspond to the co-rotation direction  ($\rm{\bf\hat{c}}$).
The top panel shows parallel (\mbox{$\rm{\bf\hat{s} \cdot \hat{c}}$ = cos$\Phi$}, the red solid line) and perpendicular (\mbox{$\rm{\bf | \hat{s}\times\hat{c} |}$ = sin$\Phi$}, the blue dashed line) components of the drift velocity at every point on the line of sight.
Here $\Phi$ is the angle between the two directions.
Small values of $\Phi$ mean the subpulses are moving along the line of sight -- this produces large phase variations in the observed drifting pattern.
At large values of $\Phi$, on the other hand, subpulses move across the line of sight. This is observed as phase-stationary intensity modulation.

\section{Results}
\label{sec:results}

The duty cycle -- the ratio of pulse width over period -- of PSR \J\ is high, $\sim0.3$.
This suggests a small inclination angle between magnetic and rotation axes.
Since the exact geometry is otherwise unknown, we use parameters in our calculations that result in the desired pulse width: $\alpha=7.7^{\circ}$, $\rho=11^{\circ}$, and $\beta=0^{\circ}$.
Note that such a geometry correspond to the anti-pulsar case with a net negative charge at the polar cap where the electric potential maximum at the polar cap determines the drift direction (see Section \ref{sec:theoretical_background}).

\subsection{Model performance}

\begin{figure}[ht]
  \centerline{\includegraphics[width=7cm]{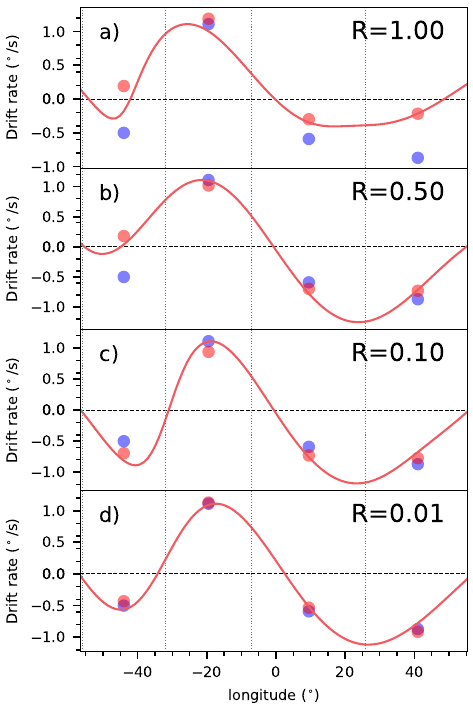}}
    \caption{
    \label{fig:4drift_characteristics}
    Modeled phase variation of subpulses (the red solid line). The blue and red dots correspond, respectively, to the measured and modeled drift rates of components. Panels show realizations with different goodness of fit (see the text for a description).}
\end{figure}

We quantify our model performance by comparing the mean phase variation of the simulated components with the measured drift rates, \mbox{$R = \sum_{n=1}^{4} \left( D_n^{\rm obs} -  D_n^{\rm sim} \right)^2$}, where  $D_n^{\rm obs}$ is the measured drift rate of the {\it n-th} component (see Table \ref{tab:drift}), and $D_n^{\rm sim}$ is the simulated mean drift rate of the {\it n-th} component.
Figure \ref{fig:4drift_characteristics} shows the modeled phase variation of subpulses with different goodness of fit. 
Panels (a) and (b) present realizations that poorly describe the drift characteristics of PSR \J~($R=1.0$, $R=0.5$), while panels (c) and (d) correspond to realizations with the observed bi-drifting characteristics ($R=0.1$, $R=0.01$). Hereafter, we define realizations as good fits if $R<0.1$.

%4drift_characteristics.pdf

% fits_2d.pdf
\begin{figure}[htb]
  \centerline{\includegraphics[width=7cm]{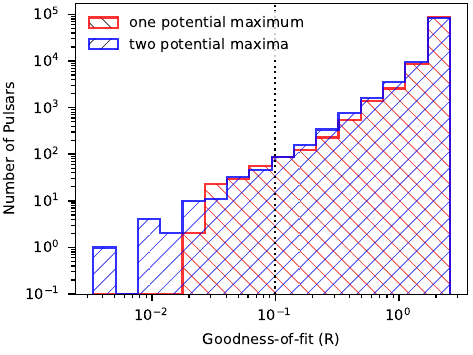}}
    \caption{
    \label{fig:fits_2d}
    Goodness of fit of realizations with dipolar magnetic field. The red and blue histograms correspond to cases with one and two potential maxima, respectively. 
    }
\end{figure}

\subsection{Dipolar magnetic field} \label{sec:dipolar}

To investigate drift in the case of a dipolar magnetic field, we consider two configurations: (I) with one potential maximum, and (II) with two potential maxima at the polar cap. 
In order to find bi-drifting, we calculated, on the Dutch national supercomputer Cartesius\footnote{\url{https://userinfo.surfsara.nl/systems/cartesius}}, the drift characteristics for 10$^5$ realizations with the same pulsar geometry, but each with random positions and amplitudes of electric potential maxima.
In Figure \ref{fig:fits_2d}, we show goodness of fit of these realizations.
The bi-drifting phenomenon is detected in 127 ($0.127 \%$) and 115 ($0.115 \%$) realizations for one and two potential maxima, respectively.

\begin{figure}[h]
  \centerline{\includegraphics[width=7cm]{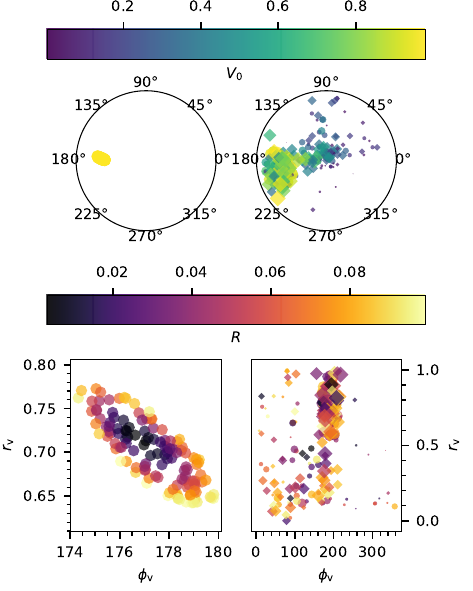}}
    \caption{
    \label{fig:vloc_2d}
    Locations of the electric potential maxima at the polar cap for the dipolar configuration of the magnetic field. The left panels correspond to the cases with a single potential maximum, while the right panels correspond to the cases with two potential maxima.
    The color bars show the potential amplitude (the upper panel) and the goodness of fit (the lower panel).
    }
\end{figure}

Figure \ref{fig:vloc_2d} shows locations of potential maxima for cases with the bi-drifting behavior.
In order to reproduce the drifting characteristics of PSR \J, a single potential maximum has to be located at \mbox{$r_{\rm v} \sim 0.7 r_{\rm pc}$}, and \mbox{$\phi_{\rm v} \sim 177^{\circ}$} (see the left panels in Figure \ref{fig:vloc_2d}), where  $r_{\rm v}$ and $\phi_{\rm v}$ are the polar coordinates of the electric potential maximum, and $r_{\rm pc}$ is the polar cap radius.
In the case of two potential maxima, the situation is less clear, but what is apparent is that the dominating potential maximum (the one with higher amplitude) is located in a similar region, \mbox{$r_{\rm v}\sim (0.6-0.9) r_{\rm pc}$}, \mbox{$\phi_{\rm v} \sim 150^{\circ}-220^{\circ}$} (see the right panels in Figure \ref{fig:vloc_2d}).

% vloc_2d.pdf

% bi_summary_6825
\begin{figure}[t]
  \centerline{\includegraphics[width=6.5cm]{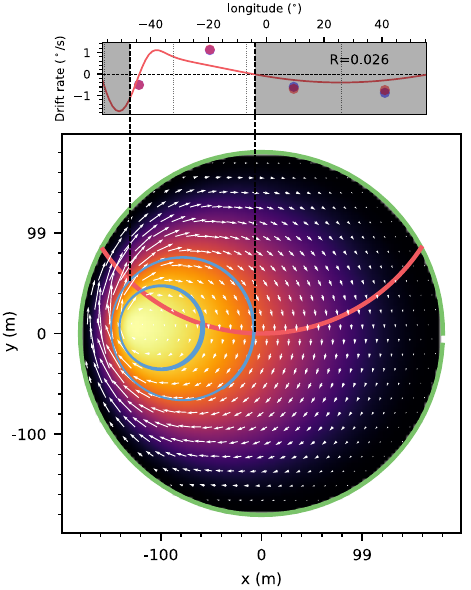}}
    \caption{
    \label{fig:bi_summary_6825}
    Drift characteristics (the top panel) and the polar cap conditions (the bottom panel) for the best realization with single potential maximum (see Figure \ref{fig:simulation} for description). 
    The blue lines in the bottom panel show spark paths at the polar cap.
     No subpulses can be seen in the grayed-out areas.
    }
\end{figure}

% bi_summary_6838
\begin{figure}[b]
  \centerline{\includegraphics[width=7cm]{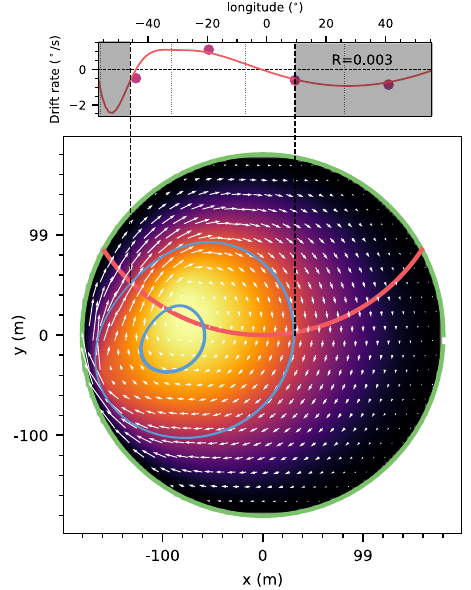}}
    \caption{
    \label{fig:bi_summary_6838}
     Drift characteristics (the top panel) and the polar cap conditions (the bottom panel) for the best realization with two potential maxima (see Figure \ref{fig:simulation} for description). 
     The blue lines in the bottom panel show spark paths at the polar cap.
     No subpulses can be seen in the grayed-out areas.
    }
\end{figure}

In Figures \ref{fig:bi_summary_6825} and \ref{fig:bi_summary_6838} we show the best models with dipolar magnetic field for single and two potential maxima, respectively. 
The models are obtained using the following parameters: $V_0=1$, $r_{\rm v}=0.72r_{\rm pc}$, and $\phi_{\rm v}=177^{\circ}$ for the single potential maximum case, and $V_0=(0.79, 0.32)$, $r_{\rm v}=(0.90 r_{\rm pc}, 0.48 r_{\rm pc})$, $\phi_{\rm v}=(194^{\circ}, 143^{\circ})$ for the case with two potential maxima.
In both cases, the modeled phase variation of the first component changes sign, which may suggest an unclear drift rate depending on the actual position of a subpulse.
This trait results in a negative drift rate for subpulses appearing in an early phase ($< -45^{\circ}$) and positive drift rate for subpulses at late phase ($> -45^{\circ}$).
However, the observed drift characteristics do not show such behavior (see the top panel in Figure \ref{fig:lrfs_data}).
To further explore if, and how, a dipolar configuration can explain the bi-drifting phenomenon, we determine the  component longitudes for the modeled configurations.
We track sparks starting on the plasma line with longitudes $-44^{\circ}$ and $-20^{\circ}$ (see the blue lines in the bottom panels in Figures \ref{fig:bi_summary_6825} and \ref{fig:bi_summary_6838}).
The resulting longitudes are [$-44^{\circ}$, $-40^{\circ}$, $-20^{\circ}$, $-3^{\circ}$], and [$-44^{\circ}$, $-30^{\circ}$, $-20^{\circ}$, $9^{\circ}$] for cases (I) and (II), respectively.
Not only is such predicted separation incompatible with the observed separation [$-44^{\circ}$, $-20^{\circ}$, $10^{\circ}$, $41^{\circ}$]; it also results in drift characteristics that are different from the one actually observed in PSR \J. 
No purely dipole magnetic field configuration was found to match the observed behavior. 

\subsection{Non-dipolar magnetic field}

\begin{figure}[h]
  \centerline{\includegraphics[width=7cm]{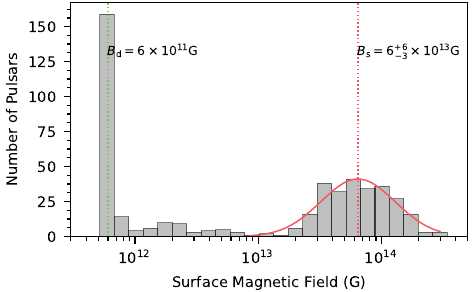}}
    \caption{
    \label{fig:field_hist}
    Surface magnetic field for realizations showing the bi-drifting behaviour ($R<0.1$).
    }
\end{figure}

To examine bi-drifting for a non-dipolar magnetic field, we calculate drift characteristics for $10^5$ realizations with one crust-anchored  magnetic anomaly with random position \mbox{$r_{\rm m,r}=0.95 R$}, \mbox{$r_{\rm m, \theta}\in (1^{\circ},70^{\circ})$}, \mbox{$r_{\rm m, \phi} \in (0^{\circ},360^{\circ})$} and strength \mbox{$B_{\rm m}\in (10^{-2}B_{\rm d}, 10^{-1}B_{\rm d})$}.
The surface magnetic field strengths that allow for bi-drifting (e.g., $R<0.1$) show an interesting dichotomy. This is clear from the histogram, Figure \ref{fig:field_hist}.
We distinguish two types of solutions: configurations with surface magnetic field of the order of dipolar component, \mbox{$B_{\rm s} \sim B_{\rm d} = 6\times 10^{11} \,{\rm G}$}, and configurations with much higher surface magnetic fields, \mbox{$B_{\rm s} \sim 10^{14} \,{\rm G}$}.
Hereafter, we refer to these cases as the low and high-field configurations.

% field_hist

\begin{figure}[ht]
  \centerline{\includegraphics[width=6.5cm]{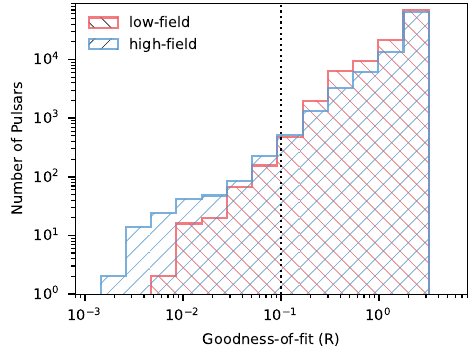}}
    \caption{
    \label{fig:goodness-of-fit}
     Goodness of fit of realizations with the non-dipolar magnetic field. The red and blue histograms correspond to cases with the low and high surface magnetic field, respectively.    
    }
\end{figure}

In the case of the non-dipolar magnetic field, the bi-drifting phenomenon is detected in 175 ($0.175 \%$) and 300 ($0.3 \%$) realizations for low and high surface magnetic field, respectively (see Figure \ref{fig:goodness-of-fit}).

% goodness-of-fit

% anomaly_locations.pdf
\begin{figure}[ht]
  \centerline{\includegraphics[width=7cm]{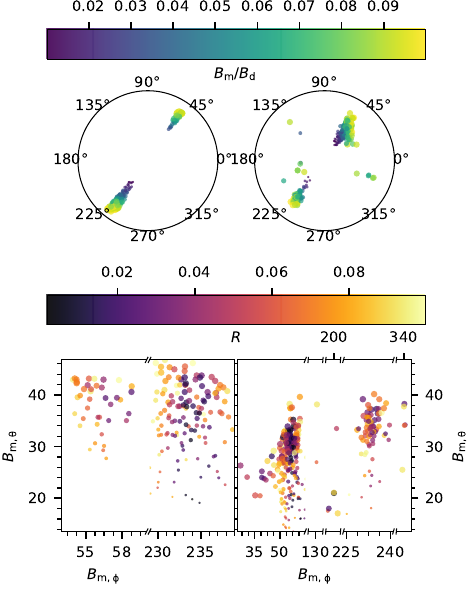}}
    \caption{
    \label{fig:anomaly_locations}
    Locations of magnetic anomalies for realizations showing bi-drifting behaviour. {\it The left panels} correspond to the low-field cases, while {\it the right panels} correspond to the high-field cases.
    The color bars show the anomaly strength ({\it the upper panel}) and the goodness-of-fit ({\it the lower panel}).
    }
\end{figure}

Figure \ref{fig:anomaly_locations} shows the locations of the magnetic anomalies for all cases with $R<0.1$.
It shows that the low-field configurations are consequence of anomalies anchored further from the magnetic axis (with polar angles \mbox{$r_{\rm m, \theta}= 38^{\circ} \pm 6^{\circ}$}) than the anomalies resulting in the high-field configurations (with \mbox{$r_{\rm m, \theta}= 29^{\circ} \pm 6^{\circ}$}).
On the other hand, the azimuthal coordinate of anomalies in both configurations are consistent, \mbox{$r_{\rm m, \phi} = 56^{\circ} \pm 4^{\circ}$} or \mbox{$r_{\rm m, \phi} =233^{\circ} \pm 3^{\circ}$}.
Note that in the high-field case the azimuthal coordinate of an anomaly for 13 realizations is beyond this range; however, the vast majority of realizations ($\sim 290$) is consistent with the low-field case.

%\mbox{$B_{\rm m, \theta}\in (1^{\circ},70^{\circ})$}, \mbox{$B_{\rm m, \phi} \in (0^{\circ},360^{\circ})$} 

\begin{figure*}[tb]
  \centerline{\includegraphics[width=16cm]{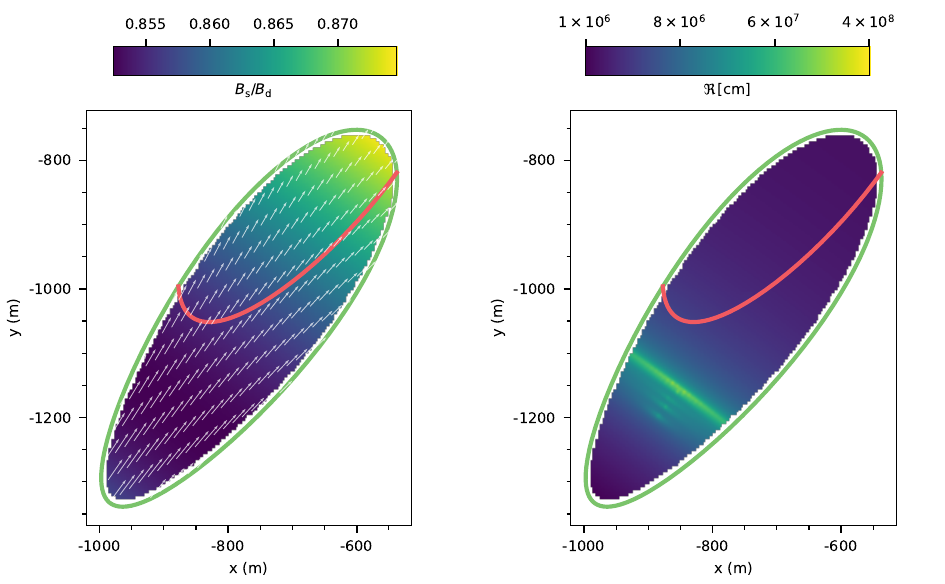}}
    \caption{
    \label{fig:bi_conditions_789}
    Magnetic field at the polar cap for the best realization with the low-field configuration. The color bar in the left panel corresponds to the magnetic field strength, while the color bar in the right panel corresponds to the radius of curvature. The white arrows in the left panel show the xy-component of the magnetic field.
    }
\end{figure*}

\begin{figure*}[tb]
  \centerline{\includegraphics[width=16cm]{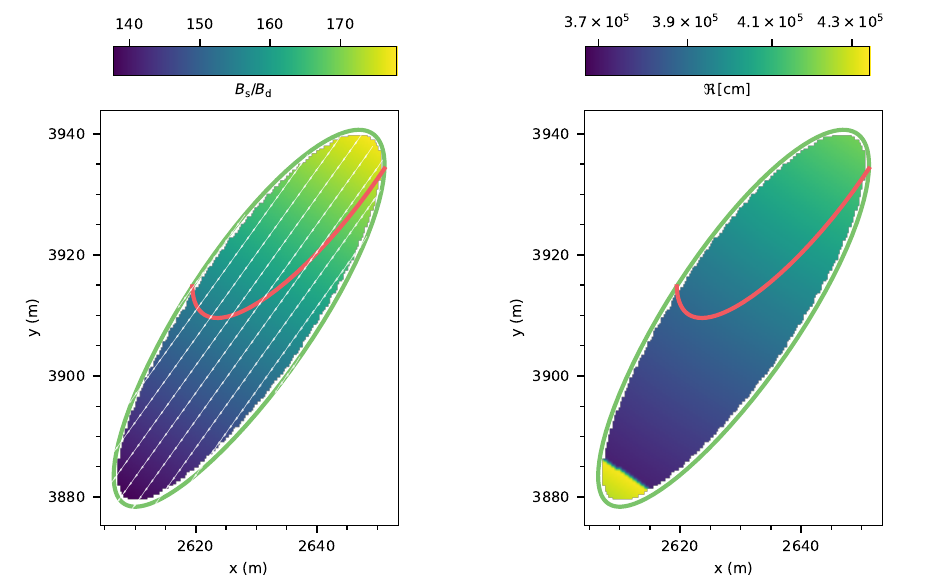}}
    \caption{
    \label{fig:bi_conditions_92}
    As Figure \ref{fig:bi_conditions_789}, but now for the high field case.
    }
\end{figure*}

\clearpage

In Figures \ref{fig:bi_conditions_789} and \ref{fig:bi_conditions_92} we show the polar-cap magnetic field for the best (lowest $R$) non-dipolar realizations, for low and high surface magnetic fields, respectively.
The surface magnetic field in the low-field case is \mbox{$B_{\rm s} \approx 0.9 B_{\rm d} = 5.4\times 10^{11} \,{\rm G}$} with radius of curvature of magnetic field lines in the plasma generation region \mbox{$\Re \sim (10^{6}$-- $10^{7}) \, {\rm cm}$}.
The high-field case, on the other hand, results in surface magnetic field \mbox{$B_{\rm s} \approx 160 B_{\rm d} = 9.6\times 10^{13} \,{\rm G}$} with a considerably smaller radius of curvature \mbox{$\Re \sim 4 \times 10^{5} \, {\rm cm}$}.

\begin{figure}[t]
  \includegraphics[width=8cm]{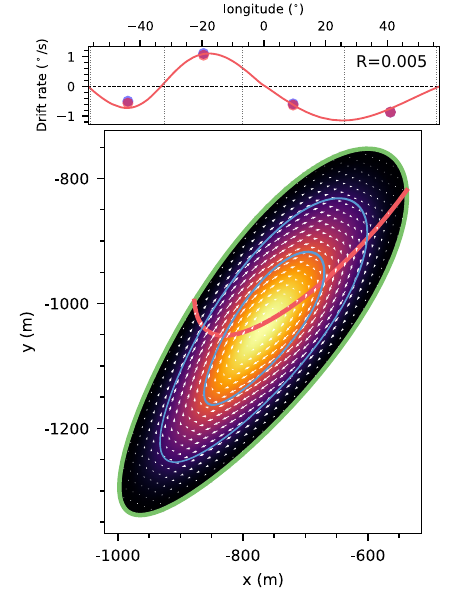}
    \caption{
    \label{fig:bi_summary_789}
    Drift characteristics (the top panel) and the polar cap conditions (the bottom panel) for the best realization with the low non-dipolar surface magnetic field (see Figure \ref{fig:simulation} for description). The blue lines in the bottom panel show spark paths at the polar cap. 
    }
\end{figure}

\begin{figure}[t]
  \includegraphics[width=8cm]{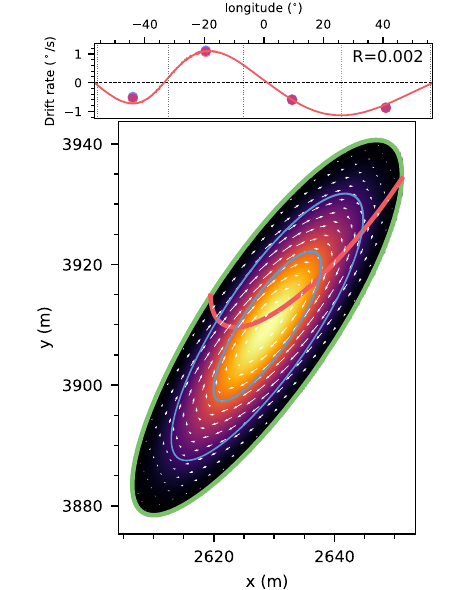}
    \caption{
    \label{fig:bi_summary_92}
    Same as Figure \ref{fig:bi_summary_789}, but now for the high-field case.
    }
\end{figure}

Figures \ref{fig:bi_summary_789} and \ref{fig:bi_summary_92}, finally, show why these two realizations are best.
The low-field configuration is a result of the magnetic anomaly with strength $B_{\rm m}=0.027 B_{\rm d}$ located at ($0.95R$, $27.86^{\circ}$, $233.58^{\circ}$).
Next, tracking the sparks  results in component longitudes [$-44^{\circ}$, $-20^{\circ}$, $19^{\circ}$, $43^{\circ}$] (see the blue lines in the bottom panel in Figure \ref{fig:bi_summary_789}). 
The high-field configuration is a result of the magnetic anomaly with strength $B_{\rm m}=0.082 B_{\rm d}$ located at ($0.95R$, $31.73^{\circ}$, $56.08^{\circ}$).
Here, the resulting component longitudes are [$-44^{\circ}$, $-20^{\circ}$, $16^{\circ}$, $42^{\circ}$] (see the blue lines in the bottom panel in Figure \ref{fig:bi_summary_92}).
Both low and high-field realizations are consistent with the observed component longitudes [$-44^{\circ}$, $-20^{\circ}$, $10^{\circ}$, $41^{\circ}$] (see Table~\ref{tab:drift}). 
Thus, non-dipolar configurations of the surface magnetic field are able to reproduce the drift and component characteristics that were observed.

% bi_summary_789 (619/789)

% bi_summary_92 (663/92)

%\subsection{Bi-drifting in J0815+0939} \label{sec:bi}

% >>> START HERE <<< 
\section{Discussion} \label{sec:discussion}

We have shown for the first time that the first component in \J~ exhibits subpulse drift in the same direction, toward earlier arrival, as components III and IV, making the diametrically opposite motion of  component II even more striking.

Such reversed motion may appear when the subpulses are aliased -- in this case, subpulses in component II would not actually be drifting at the apparent 1.1\,$\deg P^{-1}$ toward later arrival, but at much higher speed toward earlier arrival \citep[see~Figure~2 in][]{lsr+03}. 
The aliasing order of a handful of other pulsars has been determined. 
Some are shown to be unaliased (e.g. B2303+30, \citealt{2005MNRAS.357..859R}; B0809+74, \citealt{lsr+03}), while others do show the effect (e.g. B0818$-$13, \citealt{jl04}; and B1918+19, \citealt{2013MNRAS.433..445R}). One should note that in this latter category the entire subpulse system is aliased. Aliasing of only a single component has never been shown.

If component II in \J~is aliased, then all components actually drift in the same direction, yet at highly varying speeds.
While this may be conceivable in principle, we show in Figure~\ref{fig:lrfs_data} that the subpulse modulation period $P_3$ is identical for all components. 
They are thus connected through a, e.g., causal or spatial relationship. 
How one single component from such a connected set appears to move over 10 times faster than the remaining components is unclear.
In contrast with the top-right subpanel in Figure~\ref{fig:simulation}, for this situation, the drift vectors of component II would have to be closely aligned with the line-of-sight traversal, while components I, III, and IV would have to be perpendicular. 
Yet,in that case, III and IV would only show intensity variations, while phase drift is clearly seen. 
Furthermore, component II would need to span a much larger longitudinal range than I, III and IV, while this is not observed (see Table~\ref{tab:drift}).
Overall, we do not see how the reversed apparent drift of II can be the result of aliasing.

Since in \J\ only one component (II) out of four reveals drift in the opposite direction, it admittedly requires a very special setup; justified by the fact that only one such special system is known among 100 pulsars with well characterized drift properties. 

% OK. Below is some draft text, you can integrate it how you like.
Clearly the physics of drifting subpulses is not yet fully understood -- how can sparks stop moving and quench radio emission, but then after tens of seconds restart \citep[as seen for the nulls in PSR~B0809+74;][]{lsr+03}? 
Why do the sparks appear to be equidistantly spaced to such high accuracy \citep[see, e.g., the configuration proposed for PSR~B0943+10 in][]{2000_Gil}?
For these open questions, the overarching answer may be found through in determining the physical reason for the spark configuration stability in the polar cap. 
Yet even with those open questions, the carousel-like rotation of subpulses around the magnetic axis is the most viable model.

We show that such behavior is a consequence of plasma drift in the IAR around a point of electric potential extremum at the polar cap: minimum in the pulsar case ($\mathbf{\Omega} \cdot \mathbf{B} < 0$) and maximum in the antipulsar case ($\mathbf{\Omega} \cdot \mathbf{B} > 0$).
Randomly distributed discharges at the polar cap naturally results in extremum of electric potential at the center of the polar cap (see Figure \ref{fig:electric_field}).
Meanwhile, the non-dipolar configuration of the surface magnetic field next the size, shape, and location of the polar cap.
Near the surface, the drift of the plasma in the IAR is around the center of the polar cap, which can be offset from the magnetic axis.
Going up toward the the radio emission region, the magnetic field becomes ever more dipole-like.
That transition results in the drift of subpulses around the magnetic axis.

We have shown that during the discharge the plasma in a spark-forming region spins around its own axis
% rotates around a center of a spark forming region 
(see Figure \ref{fig:spark_region}). 
If isolated and alone, a spark would not drift in any particular direction.
However, the electric field between the sparks is influenced by the charge deficiency in the spark-forming regions -- and this does lead to drift around the global extremum of electric potential at the polar cap (see Figure \ref{fig:drift_velocity}).
The generally observed stability of subpulse structure suggests that the discharging regions are also stable. 
What causes this enduring existence of regions between the sparks?
One of the possible explanations is a separated flow of particles in both directions \citep[see, e.g.,][]{2003_Wright, 2012_Lyubarsky}.
Potentially the reverse flows produced in the outer magnetosphere are be responsible for maintaining a low/high (depending on pulsar geometry) intra-spark electric field.
This idea of feedback from the magnetosphere is also supported by the observations of "interpulsars", where the opposing polar caps are seen to interact \citep{1982_Fowler, 1994_Gil, 2007_Weltevrede, 2012_Weltevrede}.

We aimed to reproduce the bi-drifting phenomenon in \J~ without a strong initial preference for a model. 
We thus considered both purely dipolar as well as non-dipolar configurations of the magnetic field at the surface.
This thorough analysis over many field permutations, the first of its kind, showed that bi-drifting behavior can only be explained using a non-dipolar configuration of the surface magnetic field (see Section \ref{sec:dipolar}).
We consider that to be the main result of this work.

The multipolar nature of our best-fitting fields creates spark paths that are non-circular (see, e.g., Figure \ref{fig:bi_summary_789}). 
These resemble the elliptic, tilted spark trajectory solutions that \citet{2017_Wright} find can fit bi-drifting behavior. This agreement is assuring, and even encouraging. 
There is, however, a marked difference between the method through which these results were obtained. 
Namely, \citet{2017_Wright} geometrically map the subpulse drift onto a trajectory, while our work starts from more basic plasma principles and models a large sample of 10$^5$ possible configurations, and then naturally arrives at the observed drift behavior in a subset of cases. 
This provides better understanding of the overall landscape of (bi-)drifting phenomena; but most importantly it allows us to translate back the subpulse action into physics quantities such as magnetic field order, curvature radius, and strength.

This surface magnetic field strength in the non-dipolar case falls into one of two regions of acceptable solutions, with either a relatively low ($\sim 10^{12} \,{\rm G}$), or high ($\sim 10^{14} \,{\rm G}$) surface magnetic field.
Statistically, the strong-field solutions are more likely, as we find that $0.3\%$ realizations show bi-drifting behavior (compared to $0.175\%$ realizations resulting in relatively weak surface magnetic field).
Furthermore, it is worth mentioning that the strong surface magnetic field is consistent with predictions of the partially screened gap (see \citealt{2003_Gil, 2015_Szary}).
This high surface magnetic field is caused largely by the higher-order magnetic field components. 
The pulsar spin-down, an easily measurable quantity, will still be mostly affected by the dipolar strength of $\sim 10^{12} \,{\rm G}$, and thus cannot break the degeneracy of our solutions.

While the obtained percentages appear to be encouragingly close to the low fraction of bi-drifters known, they cannot be interpreted as the probability that bi-drifting occurs in a whole population of pulsars. 
Our calculations only covered one emission geometry.
In the underlying model, the drift characteristics are geometry dependent, and hence quantifying the overall bi-drifting probability demands population studies that more evenly sample all expected emission geometries.

However, the geometry dependence of the model can be used to put restrictions on pulsar geometry, based on drift characteristics of a pulsar alone. 
A population sample will quantify this more. 
Yet, qualitatively, we have here already shown that a bi-drifting profile such as that seen in \J\ can only be produced in a nearly aligned pulsar.

\section{Conclusions} \label{sec:conclusions}
We were able to explain the drift properties of PSR \J\ starting from a physically justified model.
We connected radio emission properties to the conditions in the IAR, and see this as a harbinger of hope for better understanding of pulsar plasma generation processes.
Finally, it opens possibilities to explore in great detail those most intriguing pulsar regions -- their polar caps. 

\acknowledgments

We would like to thank Giorgi Melikidze, Andrey Timokhin and Alexander Phillipov for constructive discussions and remarks.
Many thanks, furthermore, to the anonymous referee (you know who you are) -- your constructive comments triggered  fundamental improvements to the paper.  % TODO write it stronger
This research received funding from the Netherlands Organisation for Scientific Research (NWO) under project "CleanMachine" (614.001.301), from the European Research Council under the European Union’s Seventh Framework Programme (FP/2007-2013) / ERC Grant Agreement n. 617199, and from grant DEC-2012/05/B/ST9/03924 of the Polish National Science Centre.
Part of this work was carried out on the Dutch national e-infrastructure with the support of SURF Cooperative. Computing time was provided by NWO Physical Sciences.
The Arecibo Observatory is operated by SRI International under a cooperative agreement with the National Science Foundation (AST-1100968), and in alliance with Ana G. M\'endez-Universidad Metropolitana, and the Universities Space Research Association.

\bibliographystyle{aasjournal}

\end{document}